\documentclass[conference]{IEEEtran}
\IEEEoverridecommandlockouts

\usepackage{cite}
\usepackage{amsmath,amssymb,amsfonts}
\usepackage{algpseudocode}
\usepackage[linesnumbered,ruled]{algorithm2e}
\usepackage{array}

\newtheorem{lem}{Lemma}[section]

\usepackage{array}
\usepackage{stfloats}
\usepackage{blindtext}
\usepackage{caption}
\usepackage{subcaption}
\usepackage{textcomp}
\usepackage{stfloats}
\usepackage{url}
\usepackage{verbatim}
\usepackage{graphicx}
\usepackage{textcomp}
\usepackage{xcolor}
\def\BibTeX{{\rm B\kern-.05em{\sc i\kern-.025em b}\kern-.08em
    T\kern-.1667em\lower.7ex\hbox{E}\kern-.125emX}}
\begin{document}

\title{Demonstrating Remote Synchronization: An Experimental Approach with Nonlinear Oscillators
}
\author{\IEEEauthorblockN{Sanjeev Kumar Pandey*}
\IEEEauthorblockA{\textit{Electrical Engineering Department}
\textit{IIT Delhi}\\
Hauz Khas, New Delhi, India (110016)\\
sanjeev@ee.iitd.ac.in}
*Corresponding author
~\\
\and
\IEEEauthorblockN{Neetish Patel}
\IEEEauthorblockA{\textit{Electrical Engineering Department}
\textit{IIT Delhi}\\
Hauz Khas, New Delhi, India (110016)}
Neetish.Patel@ee.iitd.ac.in}
\maketitle

\begin{abstract}
This study investigates remote synchronization in arbitrary network clusters of coupled nonlinear oscillators, a phenomenon inspired by neural synchronization in the brain. Employing a multi-faceted approach encompassing analytical, numerical, and experimental methodologies, we leverage the Master Stability Function (MSF) to analyze network stability. We provide experimental evidence of remote synchronization between two clusters of nonlinear oscillators, where oscillators within each cluster are also remotely connected. This observation parallels the thalamus-mediated synchronization of neuronal populations in the brain. An electronic circuit testbed, supported by nonlinear ODE modeling and LT Spice simulation, was developed to validate our theoretical predictions. Future work will extend this investigation to encompass diverse network topologies and explore potential applications in neuroscience, communication networks, and power systems.
\end{abstract}


\section{Introduction}
Synchronization is common in both natural and engineered systems \cite{pikovsky2003synchronization,arenas2008synchronization,bora2024reduced,10313029}. Studies often use globally coupled oscillator models to explore coherence \cite{kuramoto1984chemical,acebron2005kuramoto,winfree1967biological}. However, natural systems typically involve heterogeneity and local coupling, which influence behaviors such as splay states and cluster synchronization \cite{pecora2014cluster,fortuna2007experimental,minati2014experimental}. Remote synchronization has promising applications \cite{olmi2024multilayer,luo2024effects,cui2024exponential,wei2024enhancing}. In neuroscience, it offers insights into how distant brain regions coordinate without direct connections, impacting cognition and information processing \cite{qin2018stability, pandey2020analyzing}. In power grids, it stabilizes dispersed generators, while in communication networks, it improves data transmission and coordination, boosting performance and reliability.

This study aims to broaden our understanding of remote synchronization by investigating its manifestation in arbitrary network clusters of coupled nonlinear oscillators. It employs a comprehensive approach encompassing analytical, numerical, and experimental methodologies. The Master Stability Function (MSF) is employed to examine the stability of the network, demonstrating analytically and computationally that a positive coupling gain results in a decrease in the Floquet multiplier and a negative MSF, ultimately leading to a stable synchronous solution. Additionally, this study highlights remote synchronization between two clusters of nonlinear oscillators, where oscillators within each cluster are remotely connected, drawing parallels to neuronal synchronization mediated by the thalamus in the brain. An electronic circuit testbed for arbitrary networks has been developed to validate these findings, supported by nonlinear ODE modeling and LT Spice simulation. This research paves the way for future investigations into synchronization phenomena within diverse network structures.
\section{Remote synchronization}
\subsection{Remote Interaction in Arbitrarily Coupled Networks}
Remote synchronization, observed in complex systems \cite{olmi2024multilayer,luo2024effects,cui2024exponential,wei2024enhancing}, describes the coordination of distinct entities without direct connections, achieved through intermediary influences. This phenomenon appears in neural networks, power grids, and social systems, revealing how collective behavior emerges without direct communication and highlighting the intricate dynamics within interconnected systems.
\begin{figure}[htp]
\centering
\includegraphics[width=0.3\textwidth]{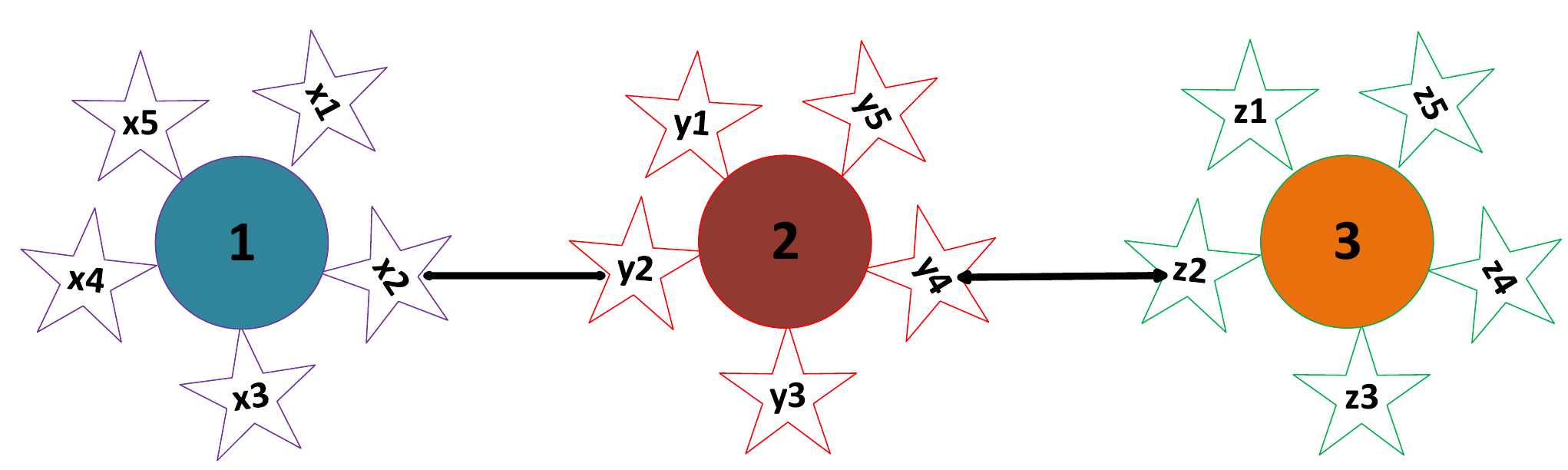}
\caption{Remotely connected network}
\label{remote_diagram}
\end{figure}

This challenges our understanding of complex systems, encouraging deeper exploration. Fig. \ref{remote_diagram} illustrates remotely connected networks, showing that nodes x1, y1, and z1 lack direct links. However, mediator nodes x2, y2, z2, and y4 enable indirect communication among clusters, allowing potential synchronization. The implications of remote synchronization are significant, providing new insights for controlling natural and engineered systems.
In neuroscience, remote synchronization reveals how distant brain regions coordinate activity without direct connections, impacting cognitive functions and information processing \cite{frank2000towards}. In power grids, it ensures stable operation of geographically dispersed generators. In communication networks, it facilitates efficient data transmission and coordination among distributed nodes, improving performance and reliability.
\subsection{Mathematical Approaches for Analyzing Remote Synchronization in Complex Systems}

\subsubsection{Lyapunov-Floquet transformation algorithm}
This section presents the Lyapunov-Floquet (Lya-Flo) transformation algorithm (\ref{1}),
\begin{algorithm}[]
\caption{Lya-FloTransformation Algorithm }
\begin{algorithmic}
\State \textbf{Input:} Periodic matrix \( A(t) \) with period \( T \).
\State \textbf{Output:} Lyapunov-Floquet transformation matrix \( P(t) \) and time-invariant matrix \( J \).
\State \textbf{Step 1: Compute the Monodromy Matrix}
\State Compute the state transition matrix \( \Phi(t) \) by solving the differential equation:
\[
\dot{\Phi}(t) = A(t)\Phi(t), \quad \Phi(0) = I_n
\]
\State Evaluate the monodromy matrix \( \Phi(T) \) at \( t = T \).
\State \textbf{Step 2: Compute the Eigenvalues and Eigenvectors of the Monodromy Matrix}
\State Solve the eigenvalue problem:
\[
\Phi(T) v_i = \lambda_i v_i, \quad i = 1, 2, \dots, n
\]
\State Here, \( \lambda_i \) are the Floquet multipliers, and \( v_i \) are the corresponding eigenvectors.
\State \textbf{Step 3: Construct the Floquet Exponents}
\State Compute the Floquet exponents \( \mu_i \) from the Floquet multipliers:
\[
\mu_i = \frac{\log(\lambda_i)}{T}, \quad i = 1, 2, \dots, n
\]
\State \textbf{Step 4: Construct the Time-Invariant Matrix \( J \)}
\State Construct the diagonal matrix \( J \) with the Floquet exponents:
\[
J = \text{diag}(\mu_1, \mu_2, \dots, \mu_n)
\]
\State \textbf{Step 5: Construct the Lyapunov-Floquet Transformation Matrix \( P(t) \)}
\State Compute the periodic transformation matrix \( P(t) \) using the relation:
\[
P(t) = \Phi(t) e^{-Jt}
\]
where \( P(t) \) is periodic with period \( T \).
\State \textbf{Step 6: Apply the Lyapunov-Floquet Transformation}
\State Apply the transformation to the original system by defining the new state \( z(t) = P^{-1}(t)x(t) \):
\[
\dot{z}(t) = J z(t)
\]
\State The system is now time-invariant.
\State \textbf{Step 7: Analyze the Stability of the Transformed System}
\State Analyze the stability by examining the eigenvalues of \(J\). The system is stable if all the real parts of the Floquet exponents \( \mu_i \) are negative.
\end{algorithmic}
\label{1}
\end{algorithm}
which converts periodic linear systems into time-invariant systems for stability analysis \cite{pandey2024synchronization}. The algorithm is outlined step by step, detailing the necessary mathematical processes. 
\subsubsection{Master Stability Function}
The Master Stability Function (MSF) is a powerful tool used to analyze the stability of synchronized states in networks of coupled dynamical systems \cite{pecora1998master,aristides2024master}. It provides a general framework for determining synchronization stability by decoupling the stability problem into mode-dependent subproblems, each associated with an eigenvalue of the network's Laplacian matrix. The stability of the synchronized state can be examined across coupling strengths by measuring the MSF for distinct eigenmodes. This method allows for a unified stability analysis independent of the network structure but depends on the nature of the individual system dynamics and the coupling function.
\begin{lem}
\label{lemma:msf_stability}
Consider a network of \( N \) coupled identical dynamical systems with individual node dynamics governed by the equation
\begin{equation}\label{coupled_eq}
 \dot{\chi}_i = \psi(\chi_i) + \kappa \sum_{j=1}^{N} L_{ij} g(\chi_j),
\end{equation}
where \( \psi(\chi_i) \) represents the intrinsic node dynamics, \( g(\chi_j)=H(\chi_{j}-\chi_{i}) \) denotes the coupling function, \( \kappa \) is the coupling strength, and \( L \) is the Laplacian matrix of the network. Assume that the network is in a synchronized state such that \( \chi_1(t) = \chi_2(t) = \dots = \chi_N(t) = s(t) \) \cite{pecora1998master}.
\small
\begin{equation}\label{gamma_eq}
\dot{\eta_{i}}=[D\psi(\chi_{s})-\kappa\lambda_{i} DH\left(\chi_s\right)] \eta_{i}, \text{ } i = 1,2,...,n.
\end{equation}
$\lambda_{i}$ be the eigenvalues of the \( L \), where \( \lambda_1 = 0 \) corresponds to the synchronized mode and \( \lambda_2, \dots, \lambda_N \) represent the non-trivial eigenmodes. Define the Master Stability Function \( \Lambda(\alpha) \), where \( \alpha = -\kappa \lambda_i \) for \( i = 2, \dots, N \). Then, the synchronous state \( s(t) \) is stable iff 
\[
\Lambda(\alpha_i) < 0 \quad \text{for all non-zero modes} \quad i = 2, \dots, N.
\]
Furthermore, the network achieves stable synchronization for the coupling strengths \( \kappa \) if and only if \( \alpha_i = -\kappa \lambda_i \) lies in the region where \( \Lambda(\alpha) < 0 \) for all \( i = 2, \dots, N \).
\end{lem}
\section{Numerical simulation}
\begin{figure}[h!]
\centering
\includegraphics[width=0.3\textwidth]{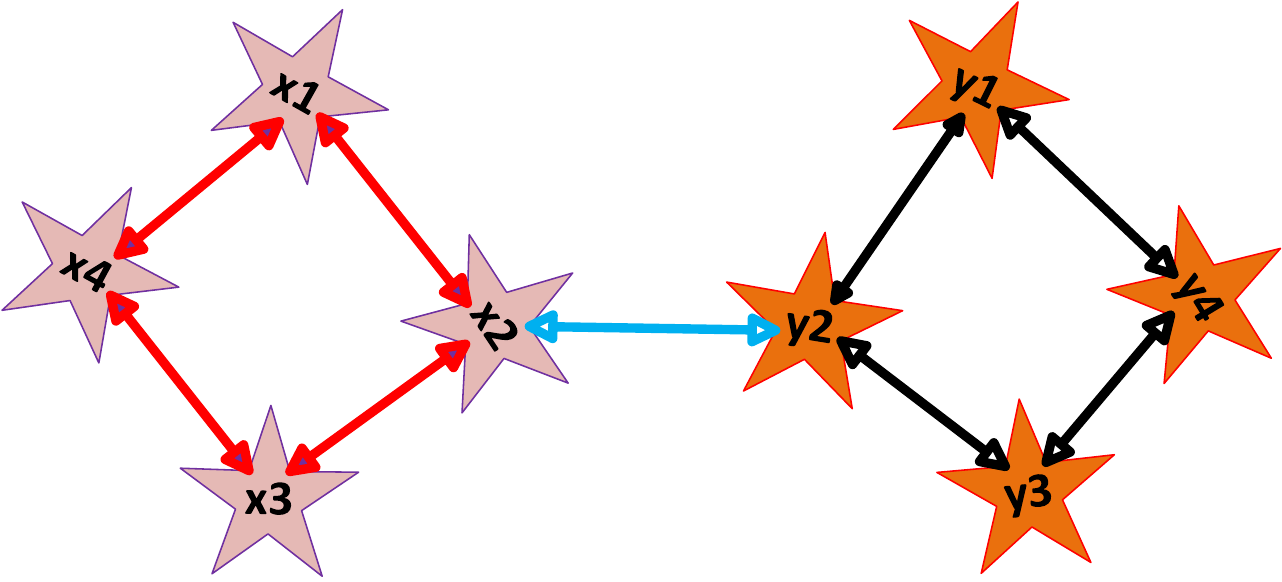}
\caption{Coupled two arbitrary network by mediaters (x2,y2).}
\label{cou_networks}
\end{figure}

The uncoupled Van der Pol oscillator (VPO) dynamics is
\small
\begin{equation} \label{eq1}
\begin{split}
\begin{array}{c}
\dot{x}_{1}=x_{2} \\
\dot{x}_{2}=\mu\left(1-x_{1}^{2}\right) x_{2}-x_{1}
\end{array}
\end{split}
\end{equation}
where $\mu$ is the nonlinear damping term. The designed remotely coupled two cluster has been shown in Fig. \ref{cou_networks}. In the coupled network (x1, x2, x3, x4), self nodes x1 and x3 are not physically connected similarly x2 and x4 are also not physically connected but they are mediated by x2 and x4. Similar observation have been observed for the y1, y2, y3, and y4. Interesting phenomenon is both the networks x and y are first remotely connected itself they they connected to each other. Here, the mediator nodes are x2 and y2 such that all the corresponding nodes get synchronized. If we choose the fixed gain for the corresponding nodes then cluster formation will observed, but in this problem all the nodes and coupling are identical and connected over a undirected graph (Fig. \ref{cou_networks}). 

The dynamics of the complete arbitrary network is
\small
\begin{equation}\label{coupled_eq}
\begin{aligned}
&\dot{x}_{11}=x_{12}+\kappa(x_{21}+x_{41}-2x_{11}), \\
&\dot{x}_{12}=\mu(1-x_{11}^{2})x_{12}-x_{11}+\kappa(x_{22}+x_{42}-2x_{12}), \\
&\dot{x}_{21}=x_{22}+\kappa(y_{21}+x_{31}+x_{11}-3x_{21}), \\
&\dot{x}_{22}=\mu(1-x_{21}^{2})x_{22}-x_{21}+\kappa(y_{22}+x_{32}+x_{12}-3x_{22}), \\
&\dot{x}_{31}=x_{32}+\kappa(x_{21}+x_{41}-2x_{31}),\\
&\dot{x}_{32}=\mu(1-x_{31}^{2})x_{32}-x_{31}+\kappa(x_{22}+x_{42}-x_{32}),\\
&\dot{x}_{41}=x_{42}+\kappa(x_{31}+x_{11}-2x_{41}), \\
&\dot{x}_{42}=\mu(1-x_{41}^{2})x_{42}-x_{41}+\kappa(x_{32}+x_{12}-2x_{42}),\\
&\dot{y}_{11}=y_{12}+\kappa(y_{21}+y_{41}-2y_{11}), \\
&\dot{y}_{12}=\mu(1-y_{11}^{2})y_{12}-y_{11}+\kappa(y_{22}+y_{42}-2y_{12}), \\
&\dot{y}_{21}=y_{22}+\kappa(x_{21}+y_{31}+y_{11}-3y_{21}), \\
&\dot{y}_{22}=\mu(1-y_{21}^{2})y_{22}-y_{21}+\kappa(x_{22}+y_{32}+y_{12}-3y_{22}), \\
&\dot{y}_{31}=y_{32}+\kappa(y_{21}-y_{41}),\\
&\dot{y}_{32}=\mu(1-y_{31}^{2})y_{32}-y_{31}+\kappa(y_{22}-y_{42}),\\
&\dot{y}_{41}=y_{42}+\kappa(y_{31}+y_{11}-2y_{41}), \\
&\dot{y}_{42}=\mu(1-y_{41}^{2})y_{42}-y_{41}+\kappa(y_{32}+y_{12}-2y_{42}).
\end{aligned}
\end{equation}

\begin{figure}[htp]
\centering
\includegraphics[width=0.35\textwidth]{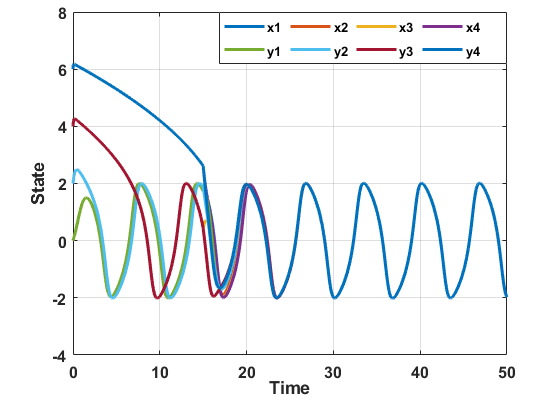}
\caption{Numerical simulation of remote synchronization for the arbitrary coupled oscillators.}
\label{ode_simulation_arbitrary}
\end{figure}

The numerical simulation of the coupled nonlinear arbitrary network has been shown in Fig. \ref{ode_simulation_arbitrary}.

\begin{figure}[htp]
\centering
\includegraphics[width=0.35\textwidth]{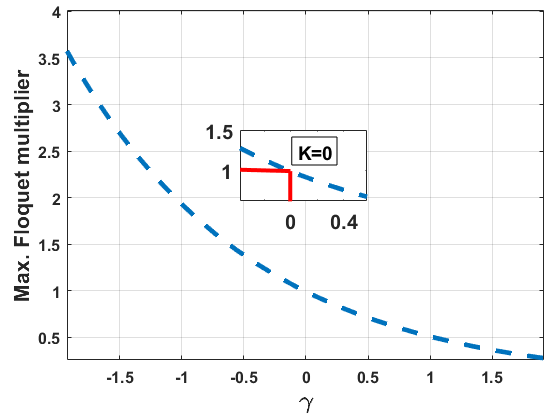}
\caption{Numerical simulation indicates that the Master Stability Function of a connected VPO exhibits a decreasing $\mu_{max}$ with increasing ($\kappa$). }
\label{msf_info_numerical}
\end{figure}

This shows the phenomenon of the remote synchronization over two coupled arbitrary network for the nonlinear oscillators. The damping parameter $\mu$ has been chosen $1$. Without coupling effect all the nodes are oscillating independently but once the coupling gain has been activated at $t=15$ sec all the nodes get synchronized. Using the graph information of both the clusters, the Laplacian matrix is given by equation \ref{eq_laplacian}. 
\small
\begin{equation} \label{eq_laplacian}
\begin{split}
\mathrm{L}=\left[\begin{array}{cccccccc}
2 & -1 & 0 & -1 & 0 & 0 & 0 & 0 \\
-1 & 3 & -1 & 0 & 0 & -1 & 0 & 0 \\
0 & -1 & 2 & -1  & 0  & 0 & 0 & 0  \\
-1 & 0 & -1 & 2 & 0  & 0 & 0 & 0\\
0 & 0 & 0 & 0 & 2 & -1 & 0 & -1\\
0 & -1 & 0 & 0 & -1 & 3 & -1 & 0\\
0 & 0 & 0 & 0 & 0 & -1 & 2 & -1\\
0 & 0 & 0 & 0 & -1 & 0 & -1 & 2
\end{array}\right]
\end{split}
\end{equation}
\normalsize
and the obtained eigenvalues from the Laplacian matrix are,
$ \lambda_{1}= 0.0$, $\lambda_{2}=0.3$,  $\lambda_{3}=2.0$, $\lambda_{4}=2.0$ , $\lambda_{5}=2.0$, $\lambda_{6}=2.8$,$\lambda_{7}=4.0$, and $\lambda_{8}=4.9$. Follows the Master Stability Function equation \ref{gamma_eq} and the algorithm (\ref{1}), the dynamics is
\small
\begin{equation*}
\begin{aligned}
&\dot{\eta}=\left[\left[\begin{array}{cc}
0 & 1 \\
-2 \mu x_{2} x_{1}-1 & \mu\left(1-x_{1}^{2}\right)
\end{array}\right] - \kappa \lambda_{max} \left[\begin{array}{cc}
1 & 0 \\
0 & 1
\end{array}\right]\right] {\eta} \\
\end{aligned}
\end{equation*}
\small \begin{equation} \label{eq5}
\begin{aligned}
&\dot{\eta}=\left[\left[\begin{array}{cc}
0 & 1 \\
-2 \mu x_{2} x_{1}-1 & \mu\left(1-x_{1}^{2}\right)
\end{array}\right] + \gamma \left[\begin{array}{cc}
1 & 0 \\
0 & 1
\end{array}\right]\right] {\eta} \\
\end{aligned}
\end{equation}

and based on (\ref{eq5}), the numerical computation of MSF has been done (Fig. \ref{msf_info_numerical}), which shows that the maximum Floquet multiplier is decreasing for $\kappa\lambda_{max}$. Here, $\lambda_{max}$ represents the maximum eigenvalues of the Laplacian matrix. Note one important point at $\kappa =0$, the $\gamma$ becomes zero, which shows that the system is uncoupled and the maximum Floquet multiplier is one. The negative behavior of the Floquet multiplier represents that the perturbation along the synchronous state (limit cycle) is decaying over time. This behavior results in the synchronization for both oscillator networks.
\begin{figure*}[t]
\centering
\includegraphics[width=\linewidth]{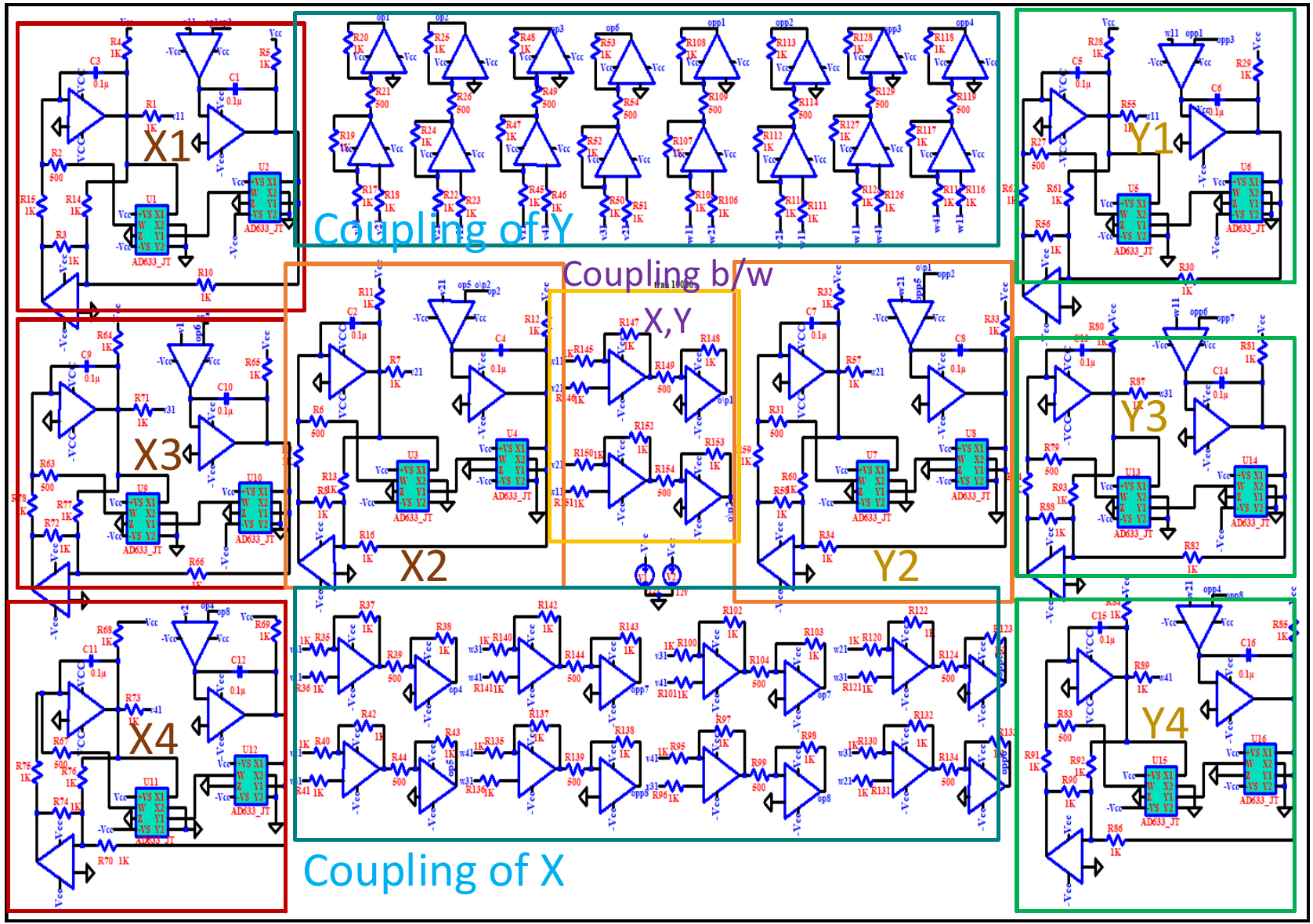}
\caption{Circuit diagram of a coupled Van der Pol oscillator network with arbitrary topology designed in LT Spice simulation.}
\label{8_node_remote_circuit}
\end{figure*}
\section{Experimental results}
The Van der Pol oscillator, a nonlinear electronic oscillator, exhibits self-sustained oscillations. It is distinguished by a negative resistance region in its operating characteristics, responsible for the energy required to sustain oscillations. The Van der Pol oscillator is a valuable tool for investigating nonlinear phenomena in various disciplines, such as electronics, physics, and biology, due to its distinctive behavior.

Following the design principles established by Roberge (1975), the oscillator circuit was implemented by utilizing resistors, operational amplifiers, analog multipliers, and capacitors. The operational amplifiers are responsible for introducing the necessary nonlinearities into the system. Concurrently, the analog multipliers facilitate the precise depiction of the quadratic elements that exist in the Van der Pol equation. The primary components ($R$ and $C$) are strategically incorporated to fine-tune the circuit's time constants, achieving the desired oscillatory behavior \cite{roberge1975operational}. The circuit diagram of the coupled two arbitrary networks is shown in Fig. \ref{8_node_remote_circuit}. 

This study investigates the synchronization behavior of two clusters of identical oscillators coupled via an arbitrary network configuration. The circuit implementation is realized in an LT Spice simulation environment first, with each network constructed utilizing components as specified in Table 1. The coupling gain, an essential factor influencing synchronization, is adjusted by altering the potentiometer's resistance. The simulation results, to be detailed later, clearly illustrate the successful synchronization of the oscillators. This experimental setup validates the theoretical framework and provides insights into the practical feasibility of achieving synchronization in coupled nonlinear systems under arbitrary network configurations.
\begin{table}[h!]
\begin{center}
\caption{Enumeration of the electronic components utilized in the experiment.}
\begin{tabular}{ | l | l | l | p{1.5cm} |}
\hline
Symbol & Parameter  & Value & Units  \\ \hline 
$R_j $ & resistor & $1.0\pm5\%$ & $M \Omega$ \\ \hline
$C_i $ & capacitor & $1.0\pm10\%$ & $ \mu F$ \\ \hline
Op-Amp  & UA741CN  &  &  \\ \hline
Potentiometer & variable resistor & $100.0\pm5\%$ & $K \Omega$ \\ \hline
$V_1$ and $V_2$ & voltage source  & $\pm 12$ & V  \\ \hline
Analog multiplier & AD633JNZ & &\\ \hline
\end{tabular}
\end{center}
\end{table}
Fig. \ref{8_node_remote_circuit} illustrates a network comprising two clusters, designated as X and Y, each consisting of four nodes. The diagram depicts that all four nodes within each cluster are interconnected through remote coupling. A similar configuration is implemented for the second cluster. Notably, within the coupled two-cluster system, nodes X2 and Y2 assume the role of mediators, facilitating indirect communication between the clusters. Upon activating the coupling gain between clusters X and Y, the system achieves synchronization, aligning both phase and frequency across all constituent oscillators. 
\begin{figure}[htp]
\centering
\includegraphics[width=0.35\textwidth]{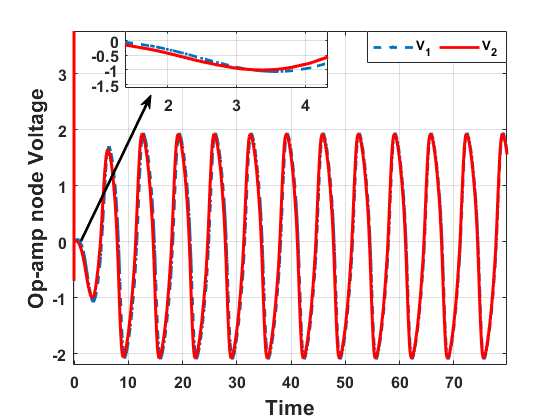}
\caption{Simulation of Coupled Van der Pol oscillators in an arbitrary network using LT Spice simulation. Here, $v_{1}$ and $v_{2}$ represents the voltage level of the both networks.}
\label{8_node_remote_LT_spice_sim}
\end{figure}
The numerical simulation conducted using LT Spice, as depicted in Fig. \ref{8_node_remote_LT_spice_sim}, clearly demonstrates the emergence of synchronization in the coupled oscillator system. Initially, the waveforms corresponding to cluster nodes X1 ($V_{1}$) and Y1 ($V_{2}$) exhibit independent oscillations characterized by distinct frequencies and phases. However, as the simulation progresses, these waveforms gradually converge, ultimately achieving synchronization in terms of both phase and frequency. This observation provides compelling evidence for the efficacy of the proposed coupling scheme in facilitating remote synchronization between the two clusters, even in the absence of direct connections between them.
The experimental validation on a breadboard, using components from Table 1, marks a significant advancement in demonstrating remote synchronization. This setup allowed precise control and monitoring of component interactions, enabling accurate synchronization assessment. The experiment not only verifies the theoretical model but also demonstrates the practical feasibility of remote synchronization in real-world conditions. This success highlights the system design's robustness and reliability, paving the way for future developments in remote synchronization technologies.
\begin{figure}[htb]
\centering
\includegraphics[width=0.4\textwidth]{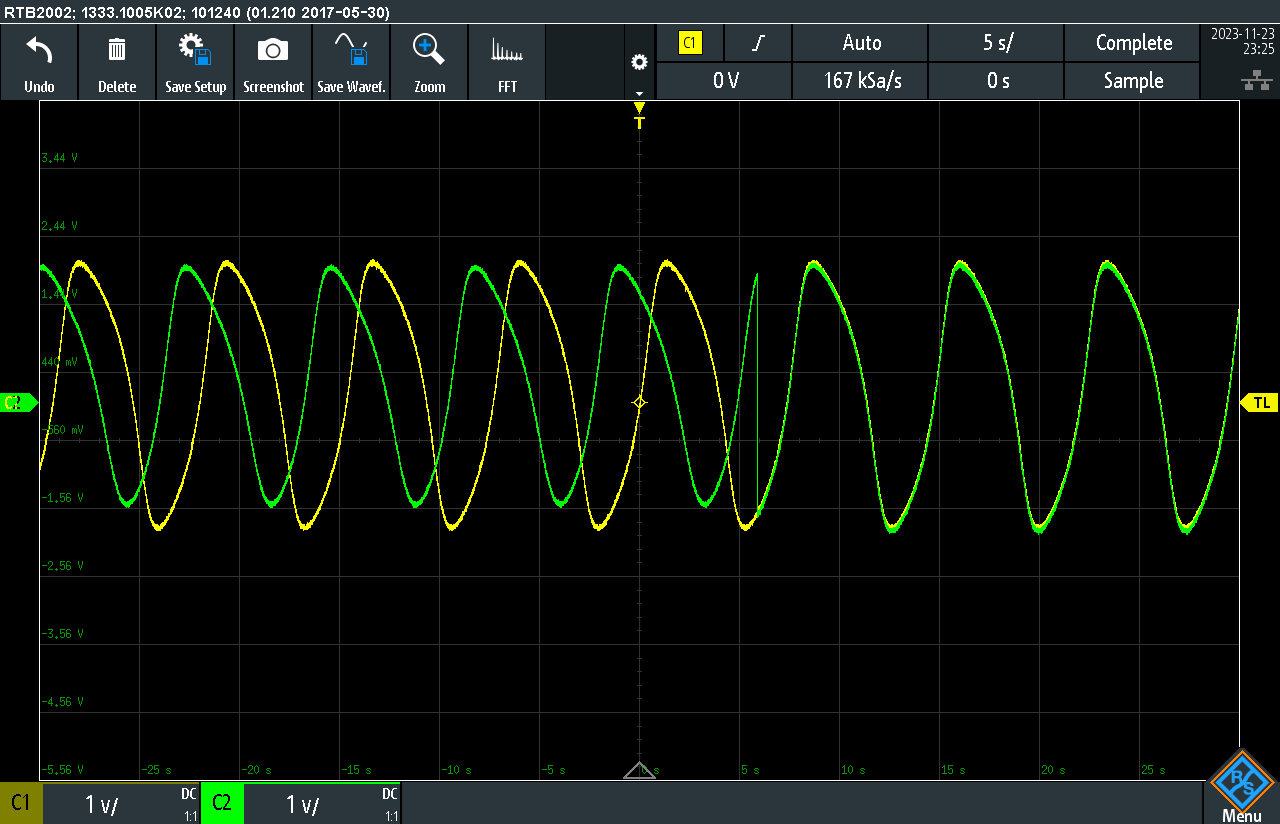}
\caption{Experimental validation of synchronization in a coupled Van der Pol oscillator network with arbitrary topology. The x-axis shows time (s), while the y-axis shows the node voltages of x1 and y1 from the two clusters.}
\label{8_node_remote_experiment}
\end{figure}
Fig. \ref{8_node_remote_experiment} shows that nodes x1 and y1, belonging to separate clusters, initially oscillate independently with a phase difference. When the coupling gain is activated by adjusting the potentiometer's resistance, both nodes synchronize, aligning their phases and frequencies. This demonstration, where nodes in distinct clusters synchronize without direct coupling, showcases the effectiveness of our research and reflects similar effects observed in brain networks and power grids, further validating our findings.

\section{Conclusion}
In conclusion, this research successfully examined remote synchronization in arbitrary clusters of coupled nonlinear oscillators through analytical, numerical, and experimental methods. The Master Stability Function provided an effective analysis of network stability, with findings highlighting the crucial role of coupling gain in achieving stable synchronous solutions, similar to neuronal synchronization in the brain. This study advances the understanding of collective behavior in complex networks, offering potential applications in neuroscience, communication systems, and power grids. Future research will explore different network topologies and advanced tools to deepen our understanding of this phenomenon.
\section*{Acknowledgments}
The author is grateful to Prof. S. Sen and Prof. I. N. Kar of the control and automation group for their unwavering support and guidance.
\bibliographystyle{ieeetr}
 \bibliography{reference}

\begin{thebibliography}{10}

\bibitem{pikovsky2003synchronization}
A.~Pikovsky, J.~Kurths, M.~Rosenblum, and J.~Kurths, {\em Synchronization: a universal concept in nonlinear sciences}.
\newblock No.~12, Cambridge university press, 2003.

\bibitem{arenas2008synchronization}
A.~Arenas, A.~D{\'\i}az-Guilera, J.~Kurths, Y.~Moreno, and C.~Zhou, ``Synchronization in complex networks,'' {\em Physics reports}, vol.~469, no.~3, pp.~93--153, 2008.

\bibitem{bora2024reduced}
R.~M. Bora and B.~B. Sharma, ``Reduced order synchronization of two non-identical special class of strict feedback systems via back-stepping control technique,'' {\em Journal of Computational and Nonlinear Dynamics}, vol.~19, no.~1, p.~011002, 2024.

\bibitem{10313029}
Y.~Qin, A.~M. Nobili, D.~S. Bassett, and F.~Pasqualetti, ``Vibrational stabilization of cluster synchronization in oscillator networks,'' {\em IEEE Open Journal of Control Systems}, vol.~2, pp.~439--453, 2023.

\bibitem{kuramoto1984chemical}
Y.~Kuramoto, ``Chemical turbulence,'' in {\em Chemical Oscillations, Waves, and Turbulence}, pp.~111--140, Springer, 1984.

\bibitem{acebron2005kuramoto}
J.~A. Acebr{\'o}n, L.~L. Bonilla, C.~J.~P. Vicente, F.~Ritort, and R.~Spigler, ``The kuramoto model: A simple paradigm for synchronization phenomena,'' {\em Reviews of modern physics}, vol.~77, no.~1, p.~137, 2005.

\bibitem{winfree1967biological}
A.~T. Winfree, ``Biological rhythms and the behavior of populations of coupled oscillators,'' {\em Journal of theoretical biology}, vol.~16, no.~1, pp.~15--42, 1967.

\bibitem{pecora2014cluster}
L.~M. Pecora, F.~Sorrentino, A.~M. Hagerstrom, T.~E. Murphy, and R.~Roy, ``Cluster synchronization and isolated desynchronization in complex networks with symmetries,'' {\em Nature communications}, vol.~5, no.~1, p.~4079, 2014.

\bibitem{fortuna2007experimental}
L.~Fortuna and M.~Frasca, ``Experimental synchronization of single-transistor-based chaotic circuits,'' {\em Chaos: An Interdisciplinary Journal of Nonlinear Science}, vol.~17, no.~4, p.~043118, 2007.

\bibitem{minati2014experimental}
L.~Minati, ``Experimental synchronization of chaos in a large ring of mutually coupled single-transistor oscillators: Phase, amplitude, and clustering effects,'' {\em Chaos: An Interdisciplinary Journal of Nonlinear Science}, vol.~24, no.~4, p.~043108, 2014.

\bibitem{olmi2024multilayer}
S.~Olmi, L.~V. Gambuzza, and M.~Frasca, ``Multilayer control of synchronization and cascading failures in power grids,'' {\em Chaos, Solitons \& Fractals}, vol.~180, p.~114412, 2024.

\bibitem{luo2024effects}
K.~Luo, Z.~Cai, Z.~Liu, S.~Guan, and Y.~Zou, ``Effects of uncommon non-isochronicities on remote synchronization,'' {\em Chaos, Solitons \& Fractals}, vol.~181, p.~114705, 2024.

\bibitem{cui2024exponential}
Y.~Cui, P.~Cheng, and X.~Ge, ``Exponential synchronization of delayed stochastic complex dynamical networks via hybrid impulsive control,'' {\em IEEE/CAA Journal of Automatica Sinica}, vol.~11, no.~3, pp.~785--787, 2024.

\bibitem{wei2024enhancing}
Z.~Wei, G.~Sriram, K.~Rajagopal, and S.~Jafari, ``Enhancing relay synchronization in multiplex networks by repulsive relay layer,'' {\em Europhysics Letters}, vol.~145, no.~2, p.~21003, 2024.

\bibitem{qin2018stability}
Y.~Qin, Y.~Kawano, and M.~Cao, ``Stability of remote synchronization in star networks of kuramoto oscillators,'' in {\em 2018 IEEE Conference on Decision and Control (CDC)}, pp.~5209--5214, IEEE, 2018.

\bibitem{pandey2020analyzing}
R.~K. Pandey and S.~K. Pandey, ``Analyzing the performance of 7nm finfet based logic circuit for the signal processing in neural network,'' in {\em 2020 IEEE Recent Advances in Intelligent Computational Systems (RAICS)}, pp.~136--140, IEEE, 2020.

\bibitem{frank2000towards}
T.~Frank, A.~Daffertshofer, C.~Peper, P.~Beek, and H.~Haken, ``Towards a comprehensive theory of brain activity:: Coupled oscillator systems under external forces,'' {\em Physica D: Nonlinear Phenomena}, vol.~144, no.~1-2, pp.~62--86, 2000.

\bibitem{pandey2024synchronization}
S.~K. Pandey, S.~Sen, and I.~N. Kar, ``Synchronization conditions for nonlinear oscillator networks,'' {\em arXiv preprint arXiv:2404.06752}, 2024.

\bibitem{pecora1998master}
L.~M. Pecora and T.~L. Carroll, ``Master stability functions for synchronized coupled systems,'' {\em Physical review letters}, vol.~80, no.~10, p.~2109, 1998.

\bibitem{aristides2024master}
R.~P. Aristides and H.~A. Cerdeira, ``Master stability functions of networks of izhikevich neurons,'' {\em Physical Review E}, vol.~109, no.~4, p.~044213, 2024.

\bibitem{roberge1975operational}
J.~K. Roberge, {\em Operational amplifiers: theory and practice}, vol.~197.
\newblock Wiley New York, 1975.

\end{thebibliography}
\end{document}